\documentclass{article}
\def\1ad{\mbox{\normalsize $^1$}}
\def\2ad{\mbox{\normalsize $^2$}}
\def\3ad{\mbox{\normalsize $^3$}}
\def\4ad{\mbox{\normalsize $^4$}}
\def\5ad{\mbox{\normalsize $^5$}}
\def\6ad{\mbox{\normalsize $^6$}}
\def\7ad{\mbox{\normalsize $^7$}}
\def\8ad{\mbox{\normalsize $^8$}}
\def\makefront{
\vspace*{1cm}\begin{center}
\def\sp{
\renewcommand{\thefootnote}{\fnsymbol{footnote}}
\footnote[4]{corresponding author : \email_speaker}
\renewcommand{\thefootnote}{\arabic{footnote}}
}
\def\newtitleline{\\ \vskip 5pt}
{\Large\bf\titleline}\\
\vskip 1truecm
{\large\bf\authors}\\
\vskip 5truemm
\addresses
\end{center}
\vskip 1truecm
{\bf Abstract:}
\abstracttext
\vskip 1truecm
}
\makeatletter
\newdimen\normalarrayskip              
\newdimen\minarrayskip                 
\normalarrayskip\baselineskip
\minarrayskip\jot
\newif\ifold             \oldtrue            \def\new{\oldfalse}
\def\arraymode{\ifold\relax\else\displaystyle\fi} 
\def\eqnumphantom{\phantom{(\theequation)}}     
\def\@arrayskip{\ifold\baselineskip\z@\lineskip\z@
     \else
     \baselineskip\minarrayskip\lineskip2\minarrayskip\fi}

\def\@arrayclassz{\ifcase \@lastchclass \@acolampacol \or
\@ampacol \or \or \or \@addamp \or
   \@acolampacol \or \@firstampfalse \@acol \fi
\edef\@preamble{\@preamble
  \ifcase \@chnum
     \hfil$\relax\arraymode\@sharp$\hfil
     \or $\relax\arraymode\@sharp$\hfil
     \or \hfil$\relax\arraymode\@sharp$\fi}}

\def\@array[#1]#2{\setbox\@arstrutbox=\hbox{\vrule
     height\arraystretch \ht\strutbox
     depth\arraystretch \dp\strutbox
     width\z@}\@mkpream{#2}\edef\@preamble{\halign \noexpand\@halignto
\bgroup \tabskip\z@ \@arstrut \@preamble \tabskip\z@ \cr}%
\let\@startpbox\@@startpbox \let\@endpbox\@@endpbox
  \if #1t\vtop \else \if#1b\vbox \else \vcenter \fi\fi
  \bgroup \let\par\relax
  \let\@sharp##\let\protect\relax
  \@arrayskip\@preamble}
\def\eqnarray{\stepcounter{equation}%
              \let\@currentlabel=\theequation
              \global\@eqnswtrue
              \global\@eqcnt\z@
              \tabskip\@centering
              \let\\=\@eqncr
              $$%
 \halign to \displaywidth\bgroup
    \eqnumphantom\@eqnsel\hskip\@centering
    $\displaystyle \tabskip\z@ {##}$%
    &\global\@eqcnt\@ne \hskip 2\arraycolsep
         $\displaystyle\arraymode{##}$\hfil
    &\global\@eqcnt\tw@ \hskip 2\arraycolsep
         $\displaystyle\tabskip\z@{##}$\hfil
         \tabskip\@centering
    &{##}\tabskip\z@\cr}
\makeatother
\def\bea{\begin{eqnarray}}
\def\eea{\end{eqnarray}}

\def\stackreb#1#2{\mathrel{\mathop{#2}\limits_{#1}}}
\def\d{\partial}
\def\beq{\begin{equation}}
\def\eeq{\end{equation}}
\def\be{\beq\new\begin{array}{c}}
\def\ee{\end{array}\eeq}                 %

\begin {document}
\def\titleline{
${\cal N}=1$ SUSY inspired Whitham prepotentials and WDVV}
\def\email_speaker{
{\tt
mironov@td.lpi.ac.ru; mironov@itep.ru
}}
\def\authors{
A.Mironov
\1ad,\2ad}
\def\addresses{
\1ad                                        %
Lebedev Physical Institute, Moscow, Russia\\            %
\2ad                                        
Institute for Theoretical and Experimental Physics\\}
\def\abstracttext{
This brief review deals with prepotentials inspired by 
${\cal N}=1$ SUSY considerations
due to Cachazo, Intrilligator, Vafa. These prepotentials associated
with matrix models following Dijkgraaf and Vafa should be considered as
given on an enlarged moduli space that includes Whitham times (couplings of
the superpotential). This moduli space is nothing but the whole
moduli space of (decorated)
hyperelliptic curves. Corresponding prepotentials are logarithms
of (quasiclassical) $\tau$-functions and satisfy the WDVV equations.}
\large
\makefront

\vspace{-12cm}

\begin{center}
\hfill FIAN/TD-04/03\\
\hfill ITEP/TH-15/03\\
\end{center}

\vspace{9.5cm}

\paragraph{\large 1. Introductory remarks.}

It was realized during last years that supersymmetric gauge theories in
the low-energy limit \cite{SW,CIV} are described by integrable structures and,
besides, can be also associated with Whitham structures
\cite{GKMMM,RG,rev,CM}. In particular, the low-energy effective action of
${\cal N}=2$ SUSY theory is described by a single function called
prepotential \cite{SW}, while the superpotential of ${\cal N}=1$ SUSY theory is
also described by a single function \cite{CIV} associated \cite{CM}
with the partition function
of Hermitian one-matrix model in a planar limit \cite{DV}. Both these
functions are associated with the $\tau$-functions of Whitham hierarchy and play a key
role in all related integrable/Whitham etc.  structures which we briefly
describe below. For the sake of brevity, we always call these 
$\tau$-functions prepotentials.

\paragraph{\large 2. Algebraic-geometrical setup of SUSY theories.}

With SUSY theories in the low-energy limit, one can associate a family of 
auxiliary Riemann surfaces so that its moduli characterize physical 
moduli/parameters of SUSY theories including v.e.v.'s (vacua) and coupling 
constants (correlation function). Both ${\cal N}=2$ and ${\cal N}=1$ SUSY 
theories are characterized by two types of variables. One set of variables, 
$\{\xi_i\}$ 
is associated with v.e.v's of various (generally composite) fields, the 
dependence on this set being usually mostly concentrated on. These variables 
correspond to A-periods of a meromorphic 1-form $dS$ given on the auxiliary 
Riemann surface. On the integrable side, these variables 
are nothing but the action variables of a proper integrable system so that 
the Jacobian of the auxiliary Riemann surface is the Liouville torus of the 
integrable system. For the genus $g$ Riemann surface there are totally $g$ 
such variables, therefore,  one deals with an integrable system with $g$ 
degrees of freedom.

The defining property of the differential $dS$ is that its variations w.r.t. 
moduli are holomorphic,
\be
{\delta dS\over\delta\hbox{moduli}}=\hbox{holomorphic}
\ee
Let us denote through $A_i$ and $B_i$ the canonical cycles on the Riemann 
surface, $A_i\circ B_j=\delta_{ij}$, through $d\omega_i$ the canonical 
holomorphic 1-differential and define the variables
\be\label{xi}
\xi_i\equiv\oint_{A_i}dS
\ee
Then one easily gets that
\be\label{ddS}
{\delta dS\over\delta \partial\xi_i}=d\omega_i
\ee
and $\oint_{B_i}dS$ is the gradient of a function $F$, 
${\partial F\over\partial\xi_i}$, since 
\be
{\partial^2 
F\over\partial\xi_i\partial\xi_j}={\partial\oint_{B_i}dS\over\partial\xi_j}=
\oint_{B_i}d\omega_j=T_{ij}
\ee
is the symmetric period matrix of the Riemann surface.

This function $F$ is exactly the function that is associated with the 
prepotential of ${\cal N}=2$ SUSY theories and the partition function of 
Hermitian one-matrix model in a planar limit. 

\paragraph{\large 3. Whitham hierarchy.}

The other set of variables, the coupling constants in SUSY gauge theories, 
$t_i$ can be associated with Whitham times on the integrable side. Indeed, 
with any solution of integrable hierarchy one can associate a Whitham 
hierarchy that describes an adiabatic evolution of moduli of the solution.
Let us consider a finite-gap solution of the hierarchy. It is 
given by an associated family of Riemann surfaces. One can 
identify this Riemann surface with the auxiliary Riemann surface.

Now the additional set of variables in terms of the auxiliary Riemann surface can be 
described as follows \cite{KriW,RG}. Let us consider a puncture, 
$P$ on the surface with a local parameter $\eta$ in its vicinity (such Riemann
surfaces with additional data are called decorated). 
Then, one can enlarge the set of 
holomorphic differentials $d\omega_i$ by the set of differentials $d\Omega_k$ 
holomorphic outside $P$ such that 
\be\label{Omega}
d\Omega_k \stackreb{\eta\to 0}{\sim} d\eta^{-k}+O(1)d\eta
\ee
This property fixes the differentials $d\Omega_k$ up to arbitrary linear 
combination of holomorphic differentials. 

Now one is ready to enlarge the set of variables by introducing times $t_k$'s,
giving the Whitham flows. The Whitham hierarchy is given by the set of equations
\be
{\partial d\omega_i\over\partial \xi_j}={\partial d\omega_j\over\partial 
\xi_i},\ \ \ \ {\partial d\Omega_k\over\partial \xi_j}=
{\partial d\omega_j\over\partial t_k},\ \ \ \ {\partial d\Omega_l\over\partial 
t_k}={\partial d\Omega_k\over\partial t_l}
\ee
These equations are solved by a 1-differential $dS$ such that
\be\label{ddS2}
{\partial dS\over\partial\xi_i}=d\omega_i,\ \ \ \ 
{\partial dS\over\partial t_k}=d\Omega_k
\ee
The first of these equations coincides with (\ref{ddS}), while the second 
one gives rise to Whitham flows. In fact, this is an equation for moduli 
dependence on Whitham variables, $t_k$. The derivatives in this equation are 
taken at constant local parameter $\eta$.\footnote{Note that this is a 
non-trivial problem to find a proper local parameter, since with most of the 
choices the Whitham equations become trivial.} The definition of $d\Omega_k$ 
and the second of equations (\ref{ddS2}) implies that
\be\label{t}
t_k={1\over k}\stackreb{\eta=0}{\hbox{res}} \left(\eta^k dS\right)
\ee
which is a counterpart of (\ref{xi}). Note that we consider $\xi_i$ and 
$t_k$ as independent variables. Therefore, one should require
\be
\oint_{A_i}d\Omega_k=\oint_{A_i}{\partial dS\over\partial t_k}=
{\partial\xi_i\over\partial t_k}=0
\ee
This normalization condition unambiguously fixes the differentials 
$d\Omega_k$.

Now one also includes the Whitham times into the set of variables of the 
function $F$,
\be\label{Ft}
{\partial F\over \partial
t_k}=\stackreb{\eta=0}{\hbox{res}} \left(\eta^{-k} dS\right)
\ee
One can check symmetricity of the proper second derivatives of $F$ using the 
Riemann identities \cite{RG} which implies that such a function $F$ does really
exists. This function is exactly the Whitham prepotential and is logarithm of a
(quasiclassical) $\tau$-function \cite{KriW,RG}. One can see that in the 
case of ${\cal N}=1$ SUSY theories this quasiclassical limit is a planar 
limit of the matrix model \cite{DV}.

\paragraph{\large 4. ${\cal N}=1$ SUSY theories.}

Now we are going to be more concrete and to consider the ${\cal N}=1$ SUSY 
theories\footnote{More details on the ${\cal N}=1$ SUSY context can be found
in \cite{DV1,DV2}.}. 
These theories are described by the superpotentials $W(\lambda)$
related to the prepotential $F$ of the following system \cite{CIV}. The 
family of Riemann surfaces is\footnote{
In fact, this 
family of Riemann surfaces is also known in ${\cal N}=2$ SUSY theories, but 
the corresponding prepotential $F$ is described by a different 
differential $dS$ \cite{rev}.}
\be\label{curve}
y^2=\left(W'(\lambda)\right)^2+f(\lambda)
\ee
i.e. these are hyperelliptic surfaces. Superpotential $W(\lambda)$ here is 
a degree $n$ polynomial, while $f(\lambda)$ is a degree $n-1$ polynomial. The 
differential 
\be\label{dS}
dS=yd\lambda
\ee
evidently gives holomorphic differentials upon varying it in moduli that 
are non-leading coefficients of the polynomial $f(\lambda)$. Therefore, one 
introduces the prepotential $F$ whose derivatives w.r.t. the A-periods of 
$dS$, $\xi_i$ are given by B-periods of $dS$. Note that there are exactly 
$n-1$ independent non-leading coefficients in $f(\lambda)$, and this is 
exactly the number of $\xi_i$'s and the genus of the Riemann surface.

Note also that in \cite{DV} it was proposed to associate with this system a 
multi-cut planar limit of the Hermitian matrix model, with the matrix model
potential coinciding with the superpotential $W(\lambda)$. Moreover, the 
partition function of this matrix model turns out to be exactly the 
prepotential $F$ \cite{CM}.

The moduli space of curves (\ref{curve}) is determined not only by the 
coefficients of $f(\lambda)$ but also by the coefficients of $W'(\lambda)$. 
Moreover, there is also the leading coefficient of $f(\lambda)$ that we
neither include into the set of variables $\xi_i$. On the other hand, the 
differential $dS$ has quite a higher order pole at $\lambda=\infty$ (there are two 
infinity points, on different sheets of the hyperelliptic Riemann surface), 
which disappears when varying in $\xi_i$. All this hints 
that one needs to extend the set of variables to 
include Whitham times, which are coefficients of the singularity of $dS$ and
are related to the coefficients of $W(\lambda)$ \cite{CM} and 
the leading coefficient of $f(\lambda)$ \cite{KM}\footnote{There is 
also another possibility, that is, to blow up the singularity at infinity 
\cite{CM}. Then, one needs to add additional handles and additional moduli 
\cite{IM1}. The extra differentials holomorphic on this higher genus surface
reduce upon degeneration to the meromorphic differentials (\ref{Omega}).
This procedure allows one to solve the problem of higher pole of 
$dS$, but one still is left with too many additional moduli of the curve that 
are not included into the set of variables.}.
Indeed, since we have a (two) puncture(s) at infinity, it is natural to choose the 
local parameter to be $\eta={1\over\lambda}$. Then, using formulas (\ref{t}), 
one realizes that the Whitham times parameterize the superpotential as 
follows\footnote{We drop the irrelevant constant term from the 
superpotential.}
\be
W(\lambda)=\sum_{k=1}^{n+1}t_k\lambda^k
\ee
The leading coefficient here can be fixed unit \cite{CMMV}.

One should also add to the set of parameters an extra time variable related 
to the leading coefficient $f_{n-1}$ of the function $f(\lambda)$ \cite{KM}
\be
t_0\equiv\ \stackreb{\infty}{\hbox{res}} \left(dS\right)={f_{n-1}\over 
2(n+1)},\ \ \ \ d\Omega_0\equiv {\partial dS\over\partial t_0}
\ee
The corresponding formula for the prepotential looks like (see \cite{CMMV})
\be
{\partial F\over\partial t_0}=\int_{-\infty}^{\infty}dS
\ee
Thus, one finally ends up with 
a generic hyperelliptic curve parameterized in a tricky way 
(\ref{curve}). It usually depends on $2n+2$ branching points, but on $2n-1$ moduli 
parameters, since one can fix any three branching points using fractional-linear
transformations. However, our data requires fixing the local parameter 
$\lambda$ which means that one can no longer use the fractional-linear 
transformation, and the dimension of this extended moduli space of 
(decorated) Riemann surfaces is equal to $2n+2$.

This is the main advantage of our approach that it effectively treats as moduli {\it 
all} the parameters emerging in the physical problem/auxiliary curve.
  
\paragraph{\large 5. WDVV equations.}

Now we come to another property that is often celebrated by Whitham 
prepotentials, that is, to the Witten-Dijkgraaf-Verlinde-Verlinde (WDVV) 
equations \cite{WDVV}. In the most general form they were written in 
\cite{MMM} as a system of algebraic relations
\be
\label{WDVV}
F_I{F}_J^{-1}F_K = F_K{F}_J^{-1}F_I, \ \ \ \ \ \ \forall\ I,J,K
\ee
for the third derivatives
\be
\label{matrF}
\|{F}_{I}\|_{JK}=
{\d^3F\over\d T_I\,\d T_J\,\d T_K} \equiv F_{IJK}
\ee
of some function $F ({\bf T})$. Have been appeared first in the context of
topological string theories \cite{WDVV}, they were rediscovered later
on in much larger class of physical theories (for a review see, e.g., 
an old survey \cite{Dub} and later papers \cite{MirWDVV}).

Within the SUSY theories framework, the WDVV equations first appeared in 
\cite{MMM}, where the ${\cal N}=2$ SUSY prepotential of many physical 
theories was proved to satisfy the WDVV equations. Recently, it was 
realized \cite{CMMV} that the ${\cal N}=1$ SUSY prepotential also satisfies 
the WDVV equations\footnote{There was realized in \cite{IM2} that 
some parts of the ${\cal N}=1$ SUSY prepotential in the case of
cubic superpotential also satisfy the WDVV equations in strange variables. 
The status of this observation is still unclear.}. The idea of proof of these 
equations \cite{MMM} is the same in both ${\cal N}=1$ and ${\cal N}=2$ SUSY cases
and is based on existence of an associative algebra of 
1-differentials. In fact, one also needs another crucial ingredient, the so 
called residue formula that expresses the third derivative of the 
prepotential via 1-differentials on the Riemann surface. This kind of 
formula was first suggested in \cite{KriW} in a quite abstract and general form 
and was later checked in concrete cases of ${\cal N}=2$ SUSY theories in 
\cite{MMM}. It was further checked in the ${\cal N}=1$ case in 
\cite{CMMV}. This latter case is new as compared with the ${\cal N}=2$ case, 
since it includes the derivatives with respect to the Whitham times, while the 
residue formula in the ${\cal N}=2$ case \cite{MMM} dealt only with the 
$\xi_i$-type variables.

More concretely, the residue formula in the ${\cal N}=1$ case \cite{CMMV}
expresses third derivatives of the prepotential via the residues at zeroes
of the differential $d\lambda$, i.e. at branching points of the 
hyperelliptic curve (\ref{curve})
\be
\label{resgen}
{\d^3 F\over \d T_I\d T_J\d T_K} = 
\stackreb{d\lambda=0}{\hbox{res}}\left(dH_IdH_JdH_K\over 
d\lambda dy\right) 
\ee
where the set of variables $\{T_I\}$ includes $\{t_k,t_0,\xi_i\}$ and the 
set of differentials $\{dH_I\}$ includes 
$\{d\Omega_k,d\Omega_0,d\omega_i\}$.

Now one should consider the algebra of 1-differentials $dH_I$ with a 
product $*$ 
\be\label{algebra}
dH_I* dH_J=C^K_{IJ}dH^K
\ee
In order to define the structure constants $C^K_{IJ}$ in this algebra,
one first needs to fix an arbitrary linear combination $dH$ of $dH_I$. Then, 
$C^K_{IJ}$ are defined by the usual (not wedge!)
product of differentials modulo $dS$
\be\label{algebra1}
dH_IdH_J=C^N_{IJ}dH_NdH+ yd\lambda d{\Re}
\ee
where $d{\Re}$ means any 1-differential. Thus defined structure constants 
definitely depend on the choice of $dH$, see \cite{MMM}. Let us choose 
$dH=dH_L$ with some $L$. 

Using the definition (\ref{algebra1}), one can either directly check that 
the algebra (\ref{algebra}) is associative, or, using hyperelliptic 
parameterization, remove the factor ${d\lambda\over y}$ in order to reduce 
the algebra to the ring of polynomials with multiplication modulo the
polynomial ideal $y^2=W'^2(\lambda)+f(\lambda)$, which is obviously
associative.

In terms of structure constants, the associativity condition can be written 
as
\be\label{ass}
C^K_{IJ}C^M_{KN}=C^K_{IN}C^M_{KJ}
\ee
Now multiplying both sides of (\ref{algebra1}) by $dH_K$, taking residues at 
$d\lambda=0$ and using the residue formula (\ref{resgen}), one obtains
\be\label{FC}
F_{IJK}=C^N_{IJ}F_{LNK}
\ee
One can easily check that the inverse matrix of $F_I$ exists, and 
substituting (\ref{FC}) into (\ref{ass}), one finally arrives at the WDVV 
equations (\ref{WDVV}).

\paragraph{\large 6. Concluding remarks.}

Note that the structures described here can be easily generalized to the case when
some of the cuts on the hyperelliptic case (\ref{curve}) shrink, i.e.,
\be
y(\lambda)=M_{n-k}(\lambda)\sqrt{\prod_{i=1}^{2k}\left(\lambda-
\mu_i\right)}\equiv M_{n-k}(\lambda)\sqrt{g_{2k}(\lambda)}
\ee
where $M_{n-k}(\lambda)$ is a polynomial of degree $n-k$ and
$g_{2k}(\lambda)$ is a polynomial of degree $2k$. This means that one is
effectively left with a new curve
\be\label{ny}
y(\lambda)=\sqrt{g_{2k}(\lambda)}
\ee
This
curve of lower genus $k-1$ along with the differential $dS=M_{n-k}(\lambda)
y(\lambda)d\lambda$ give rise to a new prepotential that depends on
$k-1$ moduli and $n+1$ Whitham times. The details of this construction can be 
found in \cite{CM}. 

{\bf Acknowledgement.} The author is grateful to L.Chekhov, A.Gorsky,
I.Krichever, A.Marshakov, A.Morozov and D.Vasiliev for useful discussions.
The work was partly supported by INTAS grant 99-0590, RFBR grant
01-02-17682a, by the Grant of Support of the Scientific
Schools 96-15-96798 and by the Volkswagen-Stiftung.

\end{document}